\documentclass[aps,showpacs,preprint]{revtex4}

\usepackage{amsmath}
\usepackage{amssymb}

\begin{document}

\title{Does the cosmological constant imply the existence of a minimum mass?}

\author{C. G. B\"ohmer}
\email{boehmer@hep.itp.tuwien.ac.at}
\affiliation{The Erwin Schr\"odinger International Institute for Mathematical
             Physics, Boltzmanngasse 9, A-1090 Wien, Austria}
\affiliation{Institut f\"ur Theoretische Physik, Technische Universit\"at Wien,
             Wiedner Hauptstr. 8-10, A-1040 Wien, Austria}

\author{T. Harko}
\email{harko@hkucc.hku.hk}
\affiliation{Department of Physics, The University of Hong Kong,
             Pokfulam Road, Hong Kong SAR, P. R. China}
\date{\today}
\preprint{Preprint ESI 1712, TUW--05--14}

\begin{abstract}
We show that in the framework of the classical general relativity
the presence of a positive cosmological constant implies the
existence of a minimal mass and of a minimal density in nature.
These results rigorously follow  from the generalized Buchdahl
inequality in the presence of a cosmological constant.
\end{abstract}

\pacs{03.70.+k, 11.90.+t, 11.10.Kk}

\maketitle


\section{Introduction}

One of the most important characteristics of compact relativistic
astrophysical objects is their maximum allowed mass. The maximum
mass is crucial for distinguishing between neutron stars and black
holes in compact binaries and in determining the outcome of many
astrophysical processes, including supernova collapse and the
merger of binary neutron stars. The theoretical values of the
maximum mass and radius for white dwarfs and neutron stars were
found by Chandrasekhar and Landau and are given by $M_{\max }\approx
\left[ \left(\hbar c/G\right)m_{B}^{-4/3}\right] ^{3/2}$ and
$R_{\max }\leq \left(\hbar /mc\right) \left( \hbar
c/Gm_{B}^{2}\right) ^{1/2}$, where $m_{B}$ is the mass of the
baryons and $m$ the mass of either electron or neutron
\cite{ShTe83}. Thus, with the exception of composition-dependent
numerical factors, the maximum mass of a degenerate star depends
only on fundamental physical constants.
 For non-rotating neutron stars with the central
pressure at their center tending to the limiting value $\rho
_{c}c^{2}$ an upper bound of around $3M_{\odot}$  has been found
\cite{Rh74}. The maximum mass of different types of astrophysical
objects (neutron stars, quark stars etc.) under different physical
conditions, including rotation and magnetic fields, was considered
by using both numerical and analytical methods (\cite{all1}, and
references therein).

With the use of the gravitational field equations for a static
equilibrium configuration, Buchdahl \cite{Bu} obtained an absolute
limit of the mass-radius ratio of a stable compact object, given
by $2GM/c^2R\leq 8/9$. This limit has been generalized in the case
of scalar-tensor theories \cite{scal}, for charged fluid spheres
\cite{char}, and for the Schwarzschild-de Sitter geometries in the
presence of a cosmological constant \cite{MaDoHa00}.

If the problem of the maximum mass of compact objects had been
considered in great detail, the  more fundamental question of the
possible existence of a minimum mass in the framework of general
relativity had been investigated at a much lesser extent. The
minimum mass of neutron stars or of white dwarfs can be
derived qualitatively from energy considerations \cite{LaLi}. A
lower limit for the radius of the neutron stars of the form $R\geq
\left(3.1125-0.44192x+2.3089x^2-0.38698x^3\right)$, with
$x=M/M_{\odot }$ and $1\leq x\leq 2.5$ has been found in
\cite{Gl00}.

At a microscopic level two basic quantities, the Planck mass $m_P$
and the Planck length $l_P$ are supposed to play a fundamental
physical role. The Planck mass is derived by equating the
gravitational radius $2Gm/c^2$ of a Schwarzschild mass with its
Compton wavelength $\hbar /mc$. The corresponding mass
$m_{Pl}=\left(c\hbar /2G\right)^{1/2}$ is of the order $m_{Pl}\approx
1.5\times 10^{-5}$ g. The Planck length is given by
$l_{Pl}=\left(\hbar G/c^3\right)^{1/2}\approx 1.6\times 10^{-33}$
cm and at about this scale quantum gravity will become important
for understanding physics. The Planck mass and length are the only
parameters with dimension mass and length, respectively, which can
be obtained from the fundamental constants $c$, $G$ and $\hbar $.

The observations of high redshift supernovae \cite{Pe99} and the
Boomerang/Maxima data \cite{Ber00}, showing that the location of
the first acoustic peak in the power spectrum of the microwave
background radiation is consistent with the inflationary
prediction $\Omega =1$, have provided compelling evidence for a
net equation of state of the cosmic fluid lying in the range
$-1\leq w=p/\rho <-1/3$. To explain these observations, two dark
components are invoked: the pressure-less cold dark matter (CDM)
and the dark energy (DE) with negative pressure. CDM contributes
$\Omega_{m}\sim 0.25$, and is mainly motivated by the theoretical
interpretation of the galactic rotation curves and large scale
structure formation. DE provides $\Omega_{DE}\sim 0.7$ and is
responsible for the acceleration of the distant type Ia
supernovae. The best candidate for the dark energy is the
cosmological constant $\Lambda$, which is usually interpreted
physically as a vacuum energy. Its size is of the order $\Lambda
\approx 3\times 10^{-56}$ cm$^{-2}$ \cite{PeRa03}. In some
theoretical models is is assumed that the cosmological constant
can be derived from the reduction to $4D$ of higher-dimensional
unified theories \cite{go98}. Since at least $70\%$ of the
Universe consists of vacuum energy, it is natural to consider
$\Lambda $ as a fundamental constant. Therefore we can chose as
the set of fundamental constants $(c,G,\hbar,\Lambda)$.

By using dimensional analysis Wesson \cite{We04} has found two
different masses which can be constructed from this set of
constants. The mass $m_P$ relevant at the quantum scale is
$m_{P}=\left(\hbar/c\right)\sqrt{\Lambda /3}
      \approx 3.5\times 10^{-66}\,{\rm g}$
while the mass $m_{PE}$ relevant to the cosmological scale is
$m_{PE}=\left(c^{2}/G\right)\sqrt{3/\Lambda }
      \approx 1\times 10^{56}\,{\rm g}$.

The interpretation of the mass $m_{PE}$ is straightforward: it is
the mass of the observable part of the universe, equivalent to
$10^{80}$ baryons of mass $10^{-24}$ g each. The interpretation of
the mass $m_P$ is more difficult. By using the dimensional
reduction from higher dimensional relativity and by assuming that
the Compton wavelength of a particle cannot take any value, Wesson
\cite{We04} proposed that the mass is quantized according to the
rule $m=(n\hbar /c)\sqrt{\Lambda /3}$. Hence $m_P$ is
the minimum mass corresponding to the ground state $n=1$.

These results about the fundamental mass have been obtained by
using a phenomenological approach. It is the purpose of the
present Letter to give a rigorous proof on the existence of a
minimum mass in general relativity. The existence of such a mass
is a direct consequence of the presence of a non-zero cosmological
constant in the gravitational field equations. Therefore these two
quantities are strongly inter-related. In order to prove the
existence of a minimum mass we follow the approach introduced by
Buchdahl \cite{Bu} and generalized to the case of a non-zero
$\Lambda $ in \cite{MaDoHa00}.

The present Letter is organized as follows. The limiting density
and mass for a general relativistic object is derived in the next
Section. We conclude our results in the last section.

\section{Lower mass and density bounds for static general relativistic spheres}

We assume that the spherically symmetric general relativistic mass
distribution is described by the metric (in the present Section we set $c=1$):
\begin{equation}
      ds^2 = -e^{\nu(r)} dt^2 + e^{\lambda(r)} dr^2
      +r^2 (d\theta^2 + \sin^2\negmedspace\theta\,d\phi^2).
\end{equation}

Static and spherically symmetric perfect fluids in general
relativity are described by three independent field equations (for
four unknown functions) that imply conservation of
energy-momentum. Eliminating the function $\nu(r)$ from the field
equations yields the well known Tolman-Oppenheimer-Volkoff
equation in the presence of a cosmological constant $\Lambda$
\cite{MaDoHa00}.

Let us introduce Buchdahl variables, defined by \cite{Bu}
\begin{equation}
      y^2 = e^{-\lambda} = 1-2w(r)r^2 -\frac{\Lambda}{3}r^2,\quad
      \zeta = e^{\nu/2},\quad x=r^2,
\end{equation}
where $w(r)$ is the mean density up to $r$, $w(r)=m(r)/r^3$ and
$m(r)$ is the mass inside radius $r$, $m(r)=4\pi \int_{0}^{r}\rho
\left(r'\right)r'^{2}dr'$, with $\rho $ the mass density of the
compact object with radius $R$.

Eliminating the pressure function from the field equations, one
obtains the following differential equation \cite{Bu,MaDoHa00}
\begin{equation}
      (y\zeta_{,x})_{,x}-\frac{1}{2}\frac{w_{,x}\zeta}{y}=0.
      \label{eq:yzeta}
\end{equation}

Eq.~(\ref{eq:yzeta}) can be used to compare solutions with
decreasing energy density with ones having constant density, for
which the second term in Eq.~(\ref{eq:yzeta}) vanishes. In the
latter case one can integrate Eq.~(\ref{eq:yzeta}) and compare it
with a decreasing solution, which then yields the generalized
Buchdahl inequality in the presence of the cosmological constant
\cite{MaDoHa00}:
\begin{equation}
      \sqrt{1-\frac{2GM}{R}-\frac{\Lambda}{3}R^2} \geq
      \frac{1}{3} - \frac{\Lambda}{12\pi G \rho}.
      \label{eq:buch1}
\end{equation}

Eq.~(\ref{eq:buch1}) provides a lower bound for the mass and
density of general relativistic objects. To prove this result, we
start by squaring Eq.~(\ref{eq:buch1}), multiplying it by $G^2
M^2$, eliminating the density on the right-hand side by
$\rho=3M/(4\pi R^3)$ and taking all terms to the left-hand side.
Then we obtain the following expression
\begin{equation}
      -2G^3 M^3 + \frac{8}{9}G^2 M^2 R - \frac{\Lambda}{3}G^2 M^2 R^3
      +\frac{2 \Lambda}{27} GM R^4 - \Bigl(\frac{\Lambda}{9}R^3\Bigr)^2 R\geq 0,
\end{equation}
which can be written as a product of three terms
\begin{equation}
      -2\left(GM+\frac{\Lambda}{6}R^3 \right)
        \left[GM-\frac{2R}{9}\left(1-\sqrt{1-\frac{3\Lambda }{4}
        R^2}\right)\right]
        \left[GM-\frac{2R}{9}\left(1+\sqrt{1-\frac{3\Lambda }{4}R^2}\right)\right] \geq 0.
        \label{eq:1}
\end{equation}

Dividing by the factor $(-2)$ reverses the inequality sign. Hence
either one or all three terms of the product must have a negative
sign in order to fulfill the latter equation.

With $\Lambda=0$ we can easily find the correct signs. The first
term is strictly positive for $\Lambda=0$ (it reads $GM$), hence
only one of the remaining terms is negative. The second term for
$\Lambda=0$ also yields $GM$. Therefore, the last term must be
negative, which for vanishing $\Lambda$ gives $2GM \leq 8R/9$,
which is nothing but the standard Buchdahl inequality \cite{Bu}.

Since the signs of the three terms are now known, let us analyze
Eq. (\ref{eq:1}) with a non-zero cosmological constant. We shall
consider separately the cases of a positive ($\Lambda >0$) and of
a negative ($\Lambda <0$) cosmological constant. For $\Lambda >0$
the analysis of the signs of Eq. (\ref{eq:1}) gives the following
algebraic
conditions to be satisfied by the mass and radius of the matter distribution and by the
cosmological constant.     \\
(i)~Positivity of the first term of~(\ref{eq:1}) implies
\begin{equation}
      GM \geq -\frac{\Lambda}{6}R^3.
\end{equation}
For positive $\Lambda$ this is trivially fulfilled.\\
(ii)~Positivity of the second term yields
\begin{equation}
      GM \geq \frac{2R}{9}\left(1-\sqrt{1-\frac{3\Lambda }{4}R^2}\right),
\end{equation}
which as before gives a lower bound on the mass. \\
(iii)~Finally, negativity of the last term of the product~(\ref{eq:1}) reads
\begin{equation}
      GM \leq \frac{2R}{9}\left(1+\sqrt{1-\frac{3\Lambda }{4}R^2}\right).
\end{equation}
\mbox{} \\
Putting the three above conditions (i)--(iii) together, leads to
\begin{equation}
      \frac{2R}{9}\left(1+\sqrt{1-\frac{3\Lambda }{4}R^2}\right) \geq GM\geq \frac{2R}{9}\left(1-\sqrt{1-\frac{3\Lambda }{4}R^2}\right), {\rm for\ } \Lambda > 0.
\end{equation}

We may Taylor expand the lower bound which then reads
$\left(2R/9\right)\left(1-\sqrt{1-3\Lambda R^2/4}\right)\approx
\Lambda R^3/12$.

Therefore for a positive cosmological constant one obtains a lower
bound for the mass and the density of a general relativistic
object, given by
\begin{equation}
      \label{minm}
      2GM \geq \frac{\Lambda }{6}R^3, \qquad \rho=\frac{3M}{4\pi R^3}
      \geq \frac{\Lambda }{16\pi G} =: \rho_{\min}, \Lambda \geq 0.
\end{equation}

In the case of a negative cosmological constant, $\Lambda <0$, by
repeating the previous analysis of the signs in Eq. (\ref{eq:1})
we obtain the condition
\begin{equation}
      \frac{2R}{9}\left(1+\sqrt{1-\frac{3\Lambda }{4}R^2}\right) \leq GM\leq \frac{2R}{9}\left(1-\sqrt{1-\frac{3\Lambda }{4}R^2}\right), {\rm for\ } \Lambda < 0.
\end{equation}

By performing a small $\Lambda $ Taylor expansion we find
\begin{equation}\label{min1}
4\frac{R}{9}-\frac{\Lambda }{12}R^3  \leq GM \leq -\frac{\Lambda
}{6} R^3, {\rm for\ } \Lambda < 0.
\end{equation}

The original Buchdahl inequality \cite{Bu}, with $\Lambda = 0$
requires that $4R/9 \geq GM$. Since $\Lambda <0$ we may write Eq.
(\ref{min1}) as
\begin{equation}\label{min2}
4\frac{R}{9}+\frac{|\Lambda |}{12}R^3  \leq GM \leq
+\frac{|\Lambda |}{6} R^3, {\rm for\ } \Lambda < 0.
\end{equation}

Eq. (\ref{min2}), derived by assuming a negative cosmological
constant, obviously violates in the limit $\Lambda \to 0$ the
Buchdahl bound. The physical consequence of this fact is that we
could have massive fluid balls which are surrounded by a horizon.

Therefore the requirement of the absence of a regular solution
contained in the horizon rules out the possibility of the
existence of a minimum bound for the mass in the presence of a
negative cosmological constant. Moreover, the right hand side of
Eq. (\ref{min2}) gives  $GM \leq |\Lambda |R^3/6$, which would
imply the un-physical condition that the numerical value of the
minimal mass derived for $\Lambda >0$ is actually the maximal
allowed mass in nature for $\Lambda >0$.

The same results on the non-existence of a minimum mass for a
negative $\Lambda $ can be obtained by considering the inequality
$2R\left(1+\sqrt{1-3\Lambda R^2/4}\right)/9 \leq GM\leq -\Lambda
R^3/6, \Lambda < 0$, which can be obtained from Eq. (\ref{eq:1}),
by assuming that the first bracket is negative, the second is
always positive (for a negative $\Lambda $) and that the last
bracket is positive.

\section{Conclusions}

For $\Lambda=0$ Eqs.~(\ref{minm}) expresses the positivity of the
total mass and of the total energy density of a compact general
relativistic matter distribution, $M\geq 0$ and $\rho \geq 0$. For
$\Lambda >0$ we conclude that no object present in classical
general relativity can have a density that is smaller than
$\rho_{\min}$. In the derivation of this result we have also
assumed that $R<2/\left(\sqrt{3}\Lambda \right)$, that is, $R$ is
smaller than the size of the event horizon.

From Eq.~(\ref{minm}) one can estimate the numerical value of the
minimal density for a positive $\Lambda $ as $\rho _{\min
}=\Lambda c^2/16\pi G=8.0\times 10^{-30}$ g cm$^{-3}$ (in the
present Section we shall restore $c$ in all equations).

By assuming that the minimum mass in nature is $m_{P}=\left( \hbar
/c\right) \sqrt{\Lambda /3}$ \cite{We04}, it follows that the
radius corresponding to $m_{P}$ is given by
\begin{equation}
      R_{P}=48^{1/6}\left( \frac{\hbar G}{c^{3}}\right)^{1/3}
      \Lambda^{-1/6}\approx 1.9\,l_{Pl}^{2/3}\,\Lambda ^{-1/6},
\end{equation}
with the numerical value $R_P=4.7\times 10^{-13}$ cm$=4.7$ fm.
Hence, the radius $R_P$ is of the same order of magnitude as the
classical radius of the electron $r_e=e^2/m_ec^2=2.81\times
10^{-13}$ cm.

Therefore from the previous analysis we conclude that in the
framework of classical general relativity the possible existence
of a minimal mass and of a minimal density in nature is strictly
related to the presence of a positive cosmological constant. On
the other hand, the positivity of $\Lambda $ is confirmed by the
cosmological observations \cite{PeRa03}.

If an absolute minimum length does exist, then, via the first of
Eqs.~(\ref{minm}), a positive cosmological constant also implies
the existence of an absolute minimum mass in nature.

\acknowledgments The authors would like to thank to the anonymous
referee for comments and suggestions which considerably improved
the manuscript. C.G.B.~wishes to thank Herbert Balasin and
Wolfgang Kummer for their useful discussions. The work of
C.G.B.~was supported by the Junior Research Fellowship of The
Erwin Schr\"odinger International Institute for Mathematical
Physics. The work of T.H.~was supported by a Seed Funding
Programme for Basic Research of the Hong Kong Government.


\begin{thebibliography}{99}

\bibitem{ShTe83} S. L. Shapiro and S. A. Teukolsky, Black Holes, White Dwarfs, and Neutron Stars, John Wiley \& Sons, New
York, 1983.

\bibitem{Rh74}C. E. Rhoades and R. Ruffini, Phys. Rev. Lett., 32,
324, 1974.

\bibitem{all1}J. M. Lattimer and M. Prakash, Phys. Repts. 333,
121, 2000; K. S. Cheng and T. Harko, Phys. Rev. D 62, 083001,
2000; T. Harko and K. S. Cheng, Astron. Astrophys., 385, 947,
2002.

\bibitem{Bu} H. A. Buchdahl, Phys. Rev. 116, 1027, 1959.

\bibitem{scal}  T. Tsuchida, G. Kawamura and K. Watanabe,  Prog. Theor. Phys., 100, 29,
1998.

\bibitem{char}  M. K. Mak, Peter N. Dobson Jr. and T. Harko, Europhys. Lett., 55,
310, 2001.

\bibitem{MaDoHa00} M. K. Mak, P. N. Dobson, Jr. and T. Harko, Mod. Phys. Lett. A15,
2153, 2000; C. G. B\"ohmer, Gen. Rel. Grav., 36, 1039, 2004; C. G.
B\"ohmer,  gr-qc/0409030, 2004; A. Balaguera-Antolinez, C. G.
B\"ohmer and M. Nowakowski, gr-qc/0409004, 2004.

\bibitem{LaLi} L. D. Landau and E. M. Lifshitz, Statistical Physics, Part 1, Oxford,
Pergamon Press, 1980.

\bibitem{Gl00} N. K. Glendenning, Phys. Rev. Lett., 85, 1150,
2004.

\bibitem{Pe99} S. Perlmutter et al., Astrophys. J., 517, 565
1998; A. G. Riess et al., Astron. J. 116, 109 1998.

\bibitem{Ber00} P. de Bernardis et al., Nature 404, 995
2000; S. Hanany et al., Astrophys. J., 545, L5 2000.

\bibitem{PeRa03} P. J. E. Peebles and B. Ratra, Rev. Mod. Phys.
75, 559 2003; T. Padmanabhan, Phys. Repts. 380, 235, 2003.

\bibitem{go98} L.-X. Li and J. R. Gott, Phys. Rev. D, 58, 103513,
1998.

\bibitem{We04}P. S. Wesson, Mod. Phys. Lett., A19, 1995, 2004.

\end{thebibliography}
\end{document}